\documentstyle[prl,aps,floats,epsf,color]{revtex}
\begin{document}
\baselineskip=12pt
\def\be{\begin{equation}}
\def\ee{\end{equation}}
\def\bea{\begin{eqnarray}}
\def\eea{\end{eqnarray}}
\def\E{{\rm e}}
\def\bearst{\begin{eqnarray*}}
\def\eearst{\end{eqnarray*}}
\def\peleven{\parbox{11cm}}
\def\peffec{\peight{\bearst\eearst}\hfill\peleven}
\def\pspace{\peight{\bearst\eearst}\hfill}
\def\ptwelve{\parbox{12cm}}
\def\peight{\parbox{8mm}}
\twocolumn[\hsize\textwidth\columnwidth\hsize\csname@twocolumnfalse\endcsname

\title
{ Analysis of Non-stationary Data for Heart-Rate Fluctuations in
Terms of Drift and Diffusion Coefficients }

\author
{F. Ghasemi$^{1}$, Muhammad Sahimi$^{2}$, J. Peinke$^{3}$ and M.
Reza Rahimi Tabar $^{4,5}$ }

\vskip 1cm

\address
{$^1$Institute for Studies in theoretical Physics and Mathematics,
P.O.Box 19395-5531,Tehran, Iran\\
$^2$Department of Chemical Engineering, University of Southern
California, Los Angeles,
California 90089-1211, USA\\
$^3$Carl von Ossietzky University, Institute of Physics, D-26111
Oldenburg, Germany\\
$^4$CNRS UMR 6202, Observatoire de la C$\hat o$te d'Azur, BP 4229,
06304 Nice Cedex 4, France\\
$^5$Department of Physics, Sharif University of Technology, P.O.
Box 11365-9161, Tehran 11365, Iran}

 \maketitle

%\date{00/07/2000}
%\maketitle
%%%%%%%%%%%%%%%%%%%%%%%%%%%%%%%%%%%%%%%%%%%%%%%%%%%%%%
%ABSTRACT
%%%%%%%%%%%%%%%%%%%%%%%%%%%%%%%%%%%%%%%%%%%%%%%%%%%%%%
\begin{abstract}
We describe a method for analyzing the stochasticity in the
non-stationary data for the beat-to-beat fluctuations in the heart
rates of healthy subjects, as well as those with congestive heart
failure. The method analyzes the returns time series of the data
as a Markov process, and computes the Markov time scale, i.e., the
time scale over which the data are a Markov process. We also
construct an effective stochastic continuum equation for the
return series. We show that the drift and diffusion coefficients,
as well as the amplitude of the returns time series for healthy
subjects are distinct from those with CHF. Thus, the method may
potentially provide a diagnostic tool for distinguishing healthy
subjects from those with congestive heart failure, as it can
distinguish small differences between the data for the two classes
of subjects in terms of well-defined and physically-motivated
quantities.


\end{abstract}

 \hspace{.3in}
\newpage

]

 \noindent{\bf Introduction}

\bigskip
Cardiac interbeat intervals fluctuate in a complex manner
\cite{Peng93,Bunde00,Bernaola01,Schulte01,Ashkenazy01,Kuusela04,Torquato98}.
Recent studies reveal that under normal conditions, beat-to-beat
fluctuations in the heart rate may display extended correlations
of the type typically exhibited by dynamical systems far from
equilibrium. It has been shown \cite{Bunde00}, for example, that
the various stages of sleep may be characterized by long-range
correlations in the heart rates, separated by a large number of
beats.

The analysis of the interbeat fluctuations in the heart rates
belong to a much broader class of many natural, as well as
man-made, phenomena that are characterized by a degree of
stochasticity. Turbulent flows, fluctuations in the stock market
prices, seismic recordings, the internet traffic, pressure
fluctuations in chemical reactors, and the surface roughness of
many materials and rock \cite{Sahimi03}, are but a few examples of
such phenomena and systems. A long standing problem has been the
development of an effective reconstruction method for such
phenomena. That is, given a set of data for certain
characteristics of such phenomena (for example, the interbeat
fluctuations in the heart rates), one would like to develop an
effective equation that can reproduce the data with an accuracy
comparable to the measured data. Although many methods have been
suggested in the past, and considerable progress has been made,
the problem remains, to a large extent, unsolved.

In many cases the stochastic process to be analyzed is {\it
non-stationary}. If the process also exhibits extended
correlations, then deducing its statistical properties by the
standard methods of analyzing such processes is very difficult.
One approach to analyze such processes was proposed by Stanley and
co-workers
\cite{Peng93,Bernaola01,Ashkenazy01,Ivanov99,Amaral,Peng95,Peng94,Ivanov98}
and others \cite{Turcott96,Lipsitz90,Kaplan91,Iyengar96,Peng99}.
They studied data for heart-rate fluctuations, for both healthy
subjects and those with congestive heart failure (CHF), in terms
of self-affine fractal distributions, such as the fractional
Brownian motion (FBM). The FBM is a non-stationary stochastic
process which induces long-range correlations, the successive
increments of which are, however, stationary and follow a Gaussian
distribution. The power spectrum of a FBM is given by, $S(f)
\propto f^{-(2H+1)}$, where $H$ is the Hurst exponent that
characterizes the type of the correlations that the data contain.
Thus, one may distinguish healthy subjects from those with CHF in
terms of the numerical value of $H$ associated with the data:
negative or antipersistent correlations for $H<1/2$, as opposed to
positive or persistent correlations for $H>1/2$. The analysis of
Stanley and co-workers indicated that there may indeed be
long-range correlations in heart-rate fluctuations data that can
be characterized by the FBM and similar fractal distributions. In
addition, the data for healthy subjects seem to be characterized
by $H<1/2$, whereas those with CHF by $H>1/2$. This was a
significant discovery over the traditional methods of analyzing
non-stationary data for heart-rate fluctuations.

However, values of the Hurst exponent $H$ associated with the two
groups of subjects are non-universal. Thus, it would, for example,
be difficult to distinguish the two groups of subjects if their
associated Hurst exponents are both close to 1/2. In addition, the
FBM is a {\it non-self-averaging} distribution, i.e., given a
fixed Hurst exponent $H$, each realization of a FBM may be
significantly different from its other realizations with the same
$H$. As a result, estimating $H$ alone and characterizing the data
by a FBM cannot enable one to predict the {\it future} trends of
the data. One may also analyze such data by the deterended
fluctuating analysis
\cite{Bunde00,Bernaola01,Schulte01,Ashkenazy01} which, in many
cases, is capable of yielding accurate and insightful information
about the nature of the data.

Recently, a novel method of analyzing stochastic processes was
introduced \cite{Jafari03,tabar06,Ghasmei05,Ghasemi205}. It was
shown that by analyzing stochastic phenomena as Markov processes
and computing their Markov time (or length) scale (that is, the
time scale over which the process can be thought of as Markov),
one may {\it reconstruct} the original process with similar
statistical properties by constructing an effective equation that
governs the process. The constructed equation helps one to
understand the nature and properties of the stochastic process.
The method utilizes a set of experimental data for a phenomenon
which contains a degree of stochasticity, and constructs a simple
equation that governs the phenomenon
\cite{Jafari03,tabar06,Ghasmei05,Ghasemi205,Friedrich97,Davoudi99,Peinke,Friedrich00}.
The method is quite general; it is capable of providing a rational
explanation for complex features of the phenomenon. More
significantly, it requires {\it no scaling} feature.

In this paper we describe a method for analyzing non-stationary
data, and then utilize it to study the interbeat fluctuations in
the heart rates. We show that the application of the method to the
analysis of interbeat fluctuations in the heart rates may
potentially lead to a novel method for distinguishing healthy
subjects from those with CHF.

The plan of this paper is as follows. In the next section, we
describe the method. We then utilize the method to analyze data
for heart-rate fluctuations in human subjects.

\bigskip
\noindent{\bf Markov Analysis of Non-Stationary Data}

\bigskip
Given a (discrete) non-stationary time series $r_i$, we introduce
a quantity $x_i$, called the {\it return} of $r_i$, defined by
\begin{equation}
x_i=\ln(r_{i+1}/r_i)\;,
\end{equation}
where $r_i$ is the value of the stochastic quantity at step $i$.
If there are long-range positive correlations in the series, then
$r_i$ and $r_{i+1}$ are close in values and, therefore, we expect
the series $x_i$ to have very small values for all $t$. For white
noise, as well as data that exhibit negative or anti-correlations,
$r_i$ and $r_{i+1}$ can be completely different and, therefore,
the time series $x_i$ will fluctuate strongly.

\begin{figure}
\epsfxsize=8truecm\epsfbox{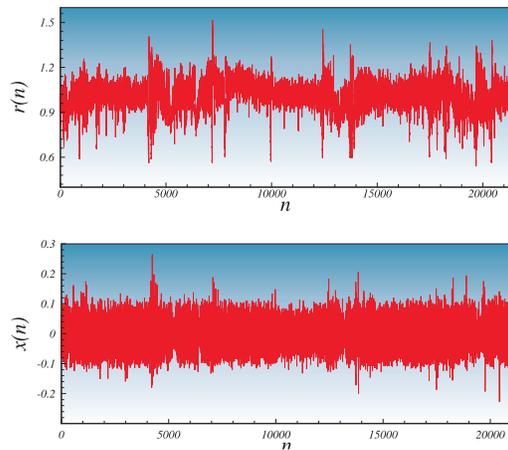}
 \narrowtext \caption{ Interbeats fluctuations of healthy subjects (top), and its returns (bottom).}
 \label{log_H}\end{figure}

\begin{figure}
\epsfxsize=8truecm\epsfbox{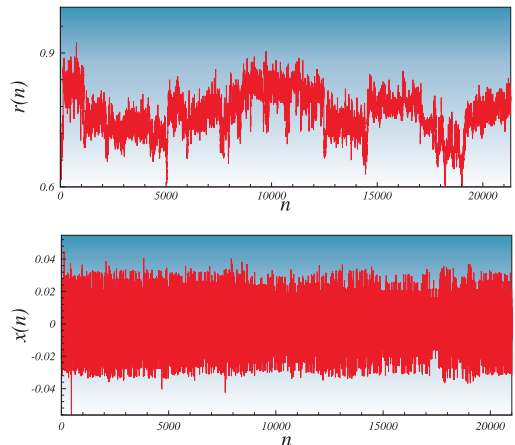}
 \narrowtext \caption{ Interbeats fluctuations of
subjects with congestive heart failure (top), and its returns
(bottom).}
 \label{log_CHF}\end{figure}

Figures \ref{log_H} and \ref{log_CHF} present the typical data
$r_i$ and the corresponding returns $x_i$ for healthy subjects and
those with CHF. The number of data is of the order of
30,000-40,000, taken over a period of about 6 hours. It is evident
that the returns series for the subjects with CHF has small
amplitudes, implying that the $r_i$ data set has long-range
positive correlations, which is consistent with the previous
analysis \cite{Peng93}. It can be verified straightforwardly that
the series $x_i$ is {\it stationary}, by measuring the stability
of its average and variance in a moving window (that is, over a
period of time which varies over the length of the series).
%%%%%%%%%%%%%%%%%%%%%%%%%%%%%%%%%%%%%%%%%%%%%%%%%%%%%%%%%%%%%%%%%%%%%%%%
%%%%%%%%%%%%%%%%%%%%%%%%%%%%%%%%%%%%%%%%%%%%%%%%%%%%%%%%%%%%%%%%%%%%%%%%
%%%%%%%%%%%%%%%%%%%%%%%%%%%%%%%%%%%%%%%%%%%%%%%%%%%%%%%%%%%%%%%%%
Due to the stationarity of the series $x(t)$, we can construct an
effective stochastic equation for the returns series of the two
groups of subjects, and distinguish the data for healthy subjects
from those with CHF. The procedure to do so involves two key
steps:

(1) Computing the Markov time scale (MTS) $t_M$ constitutes the
first step. $t_M$ is the minimum time interval over which the data
can be considered as a Markov process
\cite{Jafari03,tabar06,Ghasmei05,Ghasemi205,Siefert03}. As is
well-known, a given stochastic process with a degree of randomness
may have a finite or even an infinite $t_M$. To estimate the MTS
$t_M$, we note that a complete characterization of the statistical
properties of stochastic fluctuations of a quantity $x(t)$ in
terms of a parameter $t$ requires the evaluation of the joint
probability distribution function (PDF)
$P_n(x_1,t_1;\cdots;x_n,t_n)$ for an arbitrary $n$, the number of
the data points. If a stochastic phenomenon is a Markov process,
an important simplification can be made as $P_n$, the $n$-point
joint PDF, is generated by the product of the conditional
probabilities, $p(x_{i+1},t_{i+1}|x_i,t_i)$, for $i=1,\cdots,n-1$.

The simplest way to determine $t_M$ for stationary data is by
using the least-square test. The rigorous mathematical definition
of a Markov process is given \cite{Risken84} by
\begin{eqnarray}\label{MRR:EQ1}
&&P(x_k,t_k |x_{k-1},t_{k-1}; \cdots; x_1,t_1;x_0, t_0) \cr\nonumber\\
&& = P (x_k, t_k | x_{k-1}, t_{k-1})\;.
\end{eqnarray}
Intuitively, the physical interpretation of a Markov process is
that it "forgets its past." In other words, only the closest
"event" to $x_k$, say $x_{k-1}$ at time $t_{k-1}$, is relevant to
the probability of the event $x_k$ at $t_k$. Hence, the ability
for predicting the event $x_k$ is not enhanced by knowing its
values in steps prior to the the most recent one. Therefore, an
important simplification that is made for a Markov process is
that, the conditional multivariate joint PDF is written in terms
of the products of simple two parameter conditional PDF's
\cite{Risken84} as (\ref{MRR:EQ2})
\begin{eqnarray}\label{MRR:EQ2}
&&P(x_k,t_k;x_{k-1},t_{k-1};\cdots;x_1,t_1|x_0,t_0)\cr\nonumber \\
&&=\prod_{i=1}^k P(x_i,t_i|x_{i-1}, t_{i-1})\;.
\end{eqnarray}
Testing Eq. (\ref{MRR:EQ2}) for large values of $k$ is beyond the
current computational capability. For $k=3$ (three points or
events), however, the working equation, given by,
\begin{eqnarray}\label{MRR:EQ3}
P(x_3,t_3|x_2,t_2;x_1,t_1)=P(x_3,t_3|x_2,t_2)\;,
\end{eqnarray}
should hold for any value of $t_2$ in the interval $t_1<t_2<t_3$.
A process is then Markovian if Eq. (\ref{MRR:EQ3}) is satisfied
for a {\it certain} time separation $t_3-t_2$, in which case,
$t_M=t_3-t_2$. Thus, to compute the $t_M$ we use a fundamental
theory of probability according to which we write any three-point
PDF in terms of the conditional probability functions as,
\begin{eqnarray}\label{MRR:EQ4}
&&P(x_3,t_3;x_2,t_2;x_1,t_1)\cr \nonumber
\\ &&=P(x_3,t_3|x_2,t_2; x_1, t_1)P(x_2, t_2; x_1, t_1).
\end{eqnarray}
Using the properties of Markov processes to substitute Eq.
(\ref{MRR:EQ4}), we obtain,
\begin{eqnarray}\label{MRR:EQ5}
&& P_{\rm Markov}(x_3,t_3;x_2,t_2;x_1,t_1)\cr \nonumber \\
&& =P(x_3,t_3|x_2,t_2)P(x_2,t_2;x_1,t_1).
\end{eqnarray}
We then compare the deviation of $P_{\rm Markov}$ from that given
by Eq. (\ref{MRR:EQ4}). Using the least square method
\cite{tabar06}, we write:
\begin{eqnarray}\label{MRR:EQ6}
&& \chi^2= \int dx_3 dx_2 dx_1 \times \cr \nonumber\\
&& \frac{[P(x_3,t_3;x_2,t_2; x_1, t_1)-P_{\rm Markov}(x_3,t_3;
x_2, t_2; x_1,t_1)]^2}{\sigma^2+\sigma_{\rm Markov}^2}\;,
\end{eqnarray}
where $\sigma^2$ and $\sigma^2_{\rm Markov}$ are the corresponding
variances of terms in the nominator. Thus, one should plot the
reduced chi-square, $\chi^2_\nu=\chi^2/{\cal N}$ (with ${\cal N}$
being the number of degrees of freedom), as a function of the time
scale $t_3-t_2$. Then, $t_M=t_3-t_2$ for that value of $t_3-t_2$
for which $\chi^2_\nu$ either achieves a minimum or becomes flat
and does not change anymore; see Figure 3.

On the other hand, a necessary condition for a stochastic
phenomenon to be a Markov process is that the Chapman-Kolmogorov
(CK) equation (\ref{chap}),
\begin{eqnarray}\label{chap}
&& P(x_3,t_3|x_1,t_1)=\int\hbox{d}x_2\;P(x_3,t_3|x_2,t_2)\;P
(x_2,t_2|x_1,t_1)\;,
\end{eqnarray}
should hold for the time separation $t_3-t_2$, in which case,
$t_M=t_3-t_2$. Therefore, to test whether the time series $x(t)$
is a Matkov process, one should check the validity of the CK
equation for describing the process using different $x_1$ by
comparing the directly-evaluated conditional probability
distributions $P(x_3,t_3|x_1,t_1)$ with the one calculated
according to right side of Eq. (\ref{chap}).

(2) Estimation of the Kramers-Moyal coefficients is the second
step of constructing an effective equation for describing the
series $x_i$. The CK equation is an evolution equation for the
distribution function $P(x,t)$ at any time $t$. When formulated in
differential form, the CK equation yields the Kramers-Moyal (KM)
expansion \cite{Risken84}, given by,
\begin{eqnarray}\label{fokker1}
\frac{\partial}{\partial
t}P(x,t)=\sum_{n=1}^\infty(-\frac{\partial} {\partial x})^n
[D^{(n)}(x)P(x,t)]\;.
\end{eqnarray}
The coefficients $D^{(n)}(x,t)$ are called the KM coefficients.
They are estimated directly from the data, the conditional
probability distributions, and the moments $M^{(n)}$ defined by,
\begin{eqnarray}\label{km}
&& M^{(n)}=\frac{1}{\Delta t}\int dx'(x'-x)^n P(x',t+\Delta t |
x,t),
\cr\nonumber\\
&& D^{(n)}(x,t)=\frac{1}{n!}\hskip .2cm \lim_{\Delta t \to
0}M^{(n)}. \label{d12}
\end{eqnarray}

According to the Pawula's theorem, for a process with $D^{(4)}\sim
0$ all the $D^{(n)}$ with $n\geq 3$ vanish, in which case the KM
expansion reduces to the Fokker-Planck equation, also known as the
Kolomogrov equation \cite{Risken84}:
\begin{eqnarray}\label{fokker}
\frac{\partial}{\partial t}P(x,t)=\left[-\frac{\partial}{\partial
x}D^{(1)} (x,t)+\frac{\partial^2}{\partial
x^2}D^{(2)}(x,t)\right]P(x,t)\;.
\end{eqnarray}
Here $D^{(1)}(x,t)$ is the drift coefficient, representing the
deterministic part of the process, and $D^{(2)}(x,t)$ is the
diffusion coefficient that represents the stochastic part.

We now apply the above method to the fluctuations in the human
heartbeats of both healthy subjects and those with CHF. As
mentioned in the Introduction, several studies
\cite{Ashkenazy01,Kuusela04,tabar06,Ghasmei05,Ghasemi205,Wolf78,Ivanov99,Amaral}
indicate that, under normal conditions, the beat-to-beat
fluctuations in the heart rate may display extended correlations
of the type typically exhibited by dynamical systems far from
equilibrium, and that the two groups of subjects may be
distinguished from one another by a Hurst exponent. We show that
the drift and diffusion coefficients (as defined above) of the
interbeat fluctuations of healthy subjects and patients with CHF
have distinct behavior, when analyzed by the method we propose in
this paper, hence enabling one to distinguish the two groups of
the subjects.

We analyzed both daytime (12:00 pm to 18:00 pm) and nighttime
(12:00 am to 6:00 am) heartbeat time series of healthy subjects,
and the daytime records of patients with CHF. Our data base
includes 10 healthy subjects (7 females and 3 males with ages
between 20 and 50, and an average age of 34.3 years), and 12
subjects with CHF (3 females and 9 males with ages between 22 and
71, and an average age of 60.8 years). Figures \ref{log_H} and
\ref{log_CHF} present the data.
%%%%%%%%%%%%%%%%%%%%%%%%%%%%%%%%%%%%%%%%%%%%%%%%%%%%%%%%%%%%%%%%%%%%%%%%%
\begin{figure}
\epsfxsize=8.5truecm\epsfbox{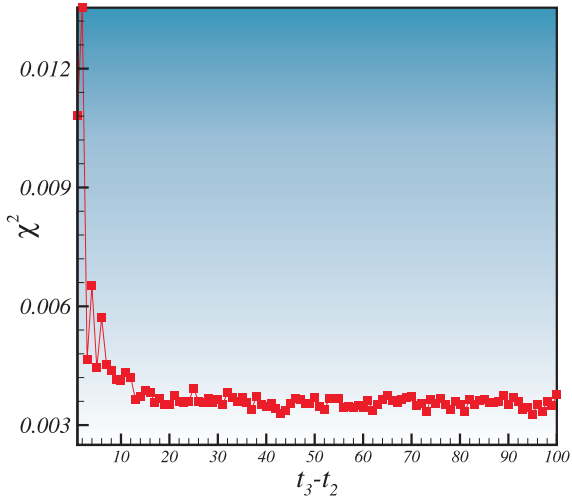} \narrowtext
\caption{$\chi^2_\nu$ values for a typical subject with CHF for
several time scales.\label{fig3}}
\end{figure}
%%%%%%%%%%%%%%%%%%%%%%%%%%%%%%%%%%%%%%%%%%%%%%%%%%%%%%%%%%%%%%%%%%%%%%%%%
We first estimate the Markov time scale $t_M$ for the returns
series of the interbeat fluctuations, using the chi-square method
described above. In Figure \ref{fig3} the results for the
$\chi^2_\nu$ values for a subject with CHF are shown. For the
healthy subjects we find the average $t_M$ for the returns, for
both the day- and nighttime data, to be (all the values are
measured in units of the average time scale for the beat-to-beat
times of each subject), $t_M=10$. On the other hand, for the
daytime records of the patients with CHF, the estimated average
$t_M$ is, $t_M=20$. Therefore, the data for the healthy subjects
are characterized by $t_M$ values that are smaller than that of
the patients with CHF by a significant factor of 2.

\begin{figure}
\epsfxsize=7truecm\epsfbox{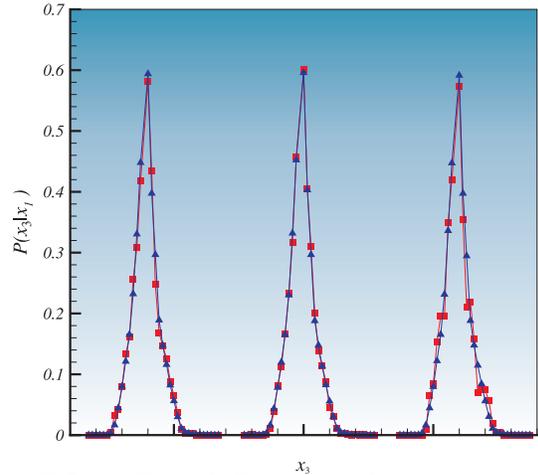} \narrowtext \caption{Test of
Chapman-Kolmogorov equation for the time separation between $t_3$
and $t_1 $ equal to the Markov time scale, for
$x_1=-6\times10^{-2}$, $x_1=0$, and $x_1=6\times10^{-2}$. Squares
and triangles represent, respectively, the directly-evaluated PDF
and that computed according to the right side of Eq. (\ref{chap}).
For clarity, the PDFs are shifted in the vertical
directions.}\label{test_chap}
\end{figure}

We then check the validity of the CK equation for several $x_1$
triplets by comparing the directly-evaluated conditional
probability distributions $P(x_3,t_3|x_1,t_1)$ with the ones
calculated according to right side of Eq. (\ref{chap}). Here, $x$
represents the returns. In Figure \ref{test_chap}, the two
differently-computed PDFs are compared. Assuming the statistical
errors to be the square root of the number of events in each bin,
we find that the two PDFs are {\it statistically identical}.
%%%%%%%%%%%%%%%%%%%%%%%%%%%%%%%%%%%%%%%%%%%%%%%%%%%%%%%%%%%%%%%%%%%%%%%%%
%%%%%%%%%%%%%%%%%%%%%%%%%%%%%%%%%%%%%%%%%%%%%%%%%%%%%%%%%%%%%%%%%

\begin{figure}
\epsfxsize=8truecm\epsfbox{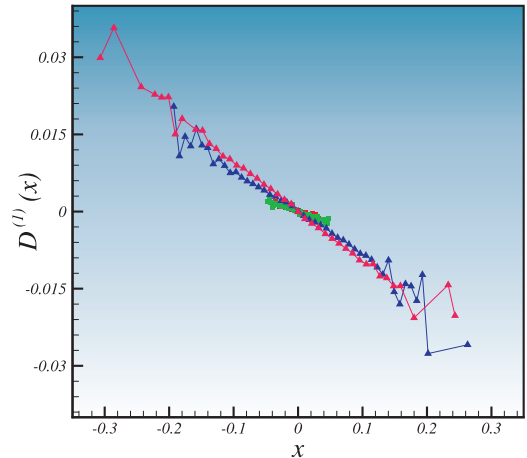}
\epsfxsize=8truecm\epsfbox{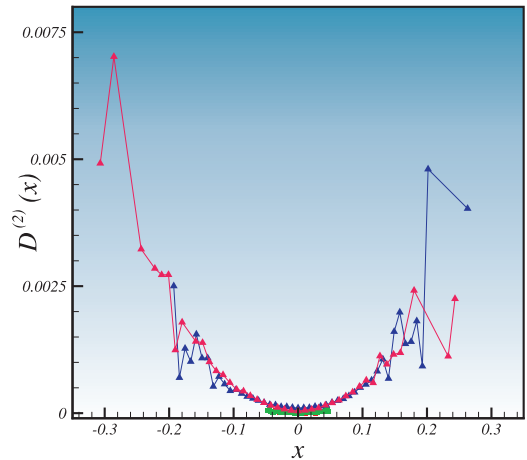} \narrowtext \caption{The
drift and diffusion coefficients, $D^{(1)}(x)$ and $D^{(2)}(x)$,
estimated by Eq. (\ref{chap}). For the healthy subjects
(triangles) and for patients with CHF (squares), $D^{(1)}(x)$ and
$D^{(2)}(x)$ follow linear and quadratic equations in $x$.}
 \label{D1-D2}\end{figure}

Using Eq. (\ref{km}) directly we calculate the drift and diffusion
coefficients, $D^{(1)}(x)$ and $D^{(2)}(x)$, for the entire set of
data for the healthy subjects, as well as those with CHF. The
corresponding $D^{(1)}(x)$ and $D^{(2)}(x)$ are displayed in
Figure \ref{D1-D2}. We find that, these coefficients provide
another important indicator for distinguishing the ill from the
healthy subjects: The drift $D^{(1)}$ and the diffusion
coefficients $D^{(2)}(x)$ follow, respectively, linear and
quadratic equations in $x$ with distinct coefficients for the
healthy subjects and patients with CHF. The analysis of the data
yields the following estimate for the healthy subjects (averaged
over the samples),
\begin{eqnarray}\label{D-H}
%& & D^{(1)}(x)= -0.11 x\;, \cr \nonumber\\
%& & D^{(2)}(x)=4\times 10 ^{-5}e-5 - 4e-5 x + 0.06 x^2\;,
& & D^{(1)}(x)= -0.1 x\;, \cr \nonumber\\
& & D^{(2)}(x)=3.7\times 10 ^{-5}- 6.6\times 10 ^{-5} x + 0.06
x^2\;,
\end{eqnarray}
with $ -0.15 < x < 0.15 $, whereas for the patients with CHF we
find that,
\begin{eqnarray}\label{D-CHF}
D^{(1)}(x)&=&-0.06 x\;, \cr \nonumber \\
D^{(2)}(x)&=&8.6\times 10 ^{-6} -2.7 \times 10 ^{-5}x + 0.03x^2\;.
\end{eqnarray}
with $ -0.04 < x < 0.04 $.

We find two important differences between the heartbeat dynamics
of the two classes of subjects:

(1) Compared with the healthy subjects, the drift and diffusion
coefficients for the patients with CHF are small.

(2) The fluctuations of the returns for healthy subjects are
distinct from those with CHF. They also fluctuate over different
intervals, indicating that the returns data for the healthy
subjects fluctuate over large interval. The fluctuations intervals
are, $-0.04<x<0.04$ and $-0.15<x<0.15$ for patients with CHF and
healthy subjects, respectively. Hence, we suggest that one may use
the drift and diffusion coefficients magnitudes, as well as the
fluctuations intervals for the returns, for characterizing the
dynamics of human heartbeats, and to distinguish healthy subjects
from those with CHF.
\section{Discussions}

Lin \cite{Lin05} argued that the daytime heart rate variability of
healthy subjects may exhibit {\it discrete} scale-invariance
(DSI). A stochastic process $x(t)$ possesses {\it continuous}
scale-invariant symmetry if its distribution is preserved under a
change of variables, $t\to\lambda t$ and $x\to x/\mu$, where
$\lambda$ and $\mu$ are {\it real} numbers, so that,
\begin{eqnarray}\label{lin}
x(t)=\frac{1}{\mu}x(\lambda t)\;.
\end{eqnarray}
If Eq.(\ref{lin}) holds only for a countable (discrete) set of
values of $\lambda$, $x(t)$ is said to possess DSI, which implies
a power-law behavior for $x(t)$ that has a log-periodic correction
of frequency $1/\log\lambda$, so that
\begin{equation}
x(t)=t^\gamma F(\log t/\log\lambda)\;,
\end{equation}
with, $\gamma=\log\mu/\log\lambda$, with $F(x)=F(x+1)$ being a
period scaling function. Generally speaking, one may write,
$x(t)=c(t)t^\zeta$, with, $\zeta= \gamma+2n\pi i/\log\lambda$,
with $n=1,2,\cdots$ The existence of log-periodicity was first
suggested by Novikov \cite{Novikov66} in small-scale energy
cascade of turbulent flows. It has been argued \cite{Sornette98}
that log-periodicity may exist in the dynamics of stock market
crashes \cite{Johansen99}, turbulence \cite{Zhou02}, earthquakes
\cite{Sornette95}, diffusion in disordered materials
\cite{Stauffer98,Saadatfar02}, and in fracture of materials near
the macroscopic fracture point \cite{Sahimi96}. The
log-periodicity, if it exists in the heart rate variability (HRV),
implies the existence of a cascade for the multifractal spectrum
of HRV, previously reported by others. However, Lin's method,
neither provides a technique for distinguishing the HRV of healthy
people from those with CHF, nor can it predict the future behavior
of HRV based on some data at earlier times.

The method proposed in the present paper is different from such
analyses in that, the {\it returns} for the data are analyzed in
terms of Markov processes. Our analysis does indicate the
existence of correlations in the return which can be quite
extended (and is characterized by the value of the Markov time
scale $t_M$).

\section{Summary}
%%%%%%%%%%%%%%%%%%%%%%%%%%%%%%%%%%%%%%%%%%%%%%%%%%%%%%%%%%%%%%%%%
We distinguish the healthy subjects from those with CHF in terms
of the {\it differences} between the drift and diffusion
coefficients of the Fokker-Plank equations that we construct for
the returns data which, in our view, provide a clearer and more
physical way of understanding the differences between the two
groups of the subjects. In addition, the reconstruction method
suggested in this paper enables one to predict the {\it future}
trends in the returns (and, hence, in the original series $r_i$)
over time scales that are of the order of the Markov time scale
$t_M$. None of the previous approaches for analyzing the data
could provide such a reconstruction method.

We also believe that, the computational method that is described
in this paper is more sensitive to small differences between the
data for healthy subjects and those with CHF. As such, it might
eventually provide a diagnostic tool for detection of CHF in
patients with small amounts of data and in its initial stages of
development.

\end{document}